\documentclass[pra,10pt,amsmath,twocolumn,floatfix]{revtex4}

\usepackage{xspace,amsmath,amsfonts,amsthm,amssymb,amsbsy,graphics,graphicx,bm,epstopdf}

\usepackage{amsfonts}
\usepackage{amssymb}
\usepackage{graphicx}
\usepackage{picture,xcolor}
\usepackage[pdftex]{hyperref}  \hypersetup{bookmarks=true}

\begin{document}
\def\sF{\mathcal{F}}
\newcommand{\bra}[1]{\left\langle{#1}\right|}
\newcommand{\ket}[1]{\left|{#1}\right\rangle}
\newcommand{\ip}[2]{\left\langle{#1}\right|\left.{#2}\right\rangle}
\newcommand{\ex}[1]{\left\langle{#1}\right\rangle}
\newcommand{\da}[1]{{#1}^{\dag}}

\newcommand{\topic}[1]{\vspace{12pt}\noindent{\bf#1}\vspace{6pt}}

\newcommand{\cmc}[1]{{\color{red}#1}}
\newcommand{\ml}[1]{{\color{blue}#1}}

\title{Optimal Quantum-Enhanced Interferometry}

\author{Matthias D.~Lang$^1$}\email{mlang@unm.edu}
\author{Carlton M.~Caves$^{1,2}$}\email{ccaves@unm.edu}

\affiliation{$^1$Center for Quantum Information and Control, University of New Mexico, Albuquerque, New Mexico, 87131-0001, USA\\
$^2$Centre for Engineered Quantum Systems, School of Mathematics and Physics, The University of Queensland, St Lucia, QLD 4072, Australia}


\begin{abstract}
We analyze the ultimate bounds on the phase sensitivity of an interferometer, given the constraint that the state input to the interferometer's initial 50:50 beam splitter $B$ is a product state of the two input modes.  Requiring a product state is a natural restriction: if one were allowed to input an arbitrary, entangled two-mode state $|\Xi\rangle$ to the beam splitter, one could generally just as easily input the state $B|\Xi\rangle$ directly into the two modes after the beam splitter, thus rendering the beam splitter unnecessary.  We find optimal states for a fixed photon number and for a fixed mean photon number.
\end{abstract}

\date{\today}

\maketitle

\section{Interferometric setting}

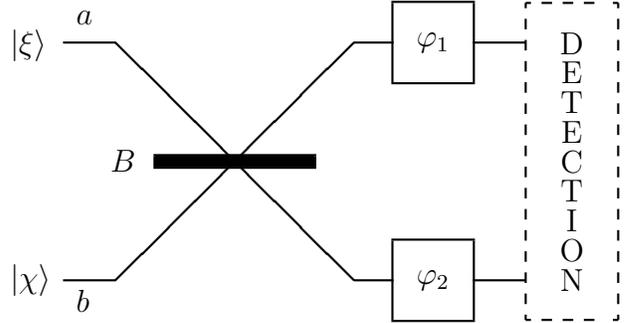
\begin{figure}[h]
\begin{center}
\begin{picture}(240,130)
\thicklines

\put(0,106){\large$|\xi \rangle$}
\put(0,17){\large$|\chi \rangle$}

\put(38,61){\large$B$}
\put(25,117){\large$a$}
\put(20,110){\line(1,0){20}}
\put(20,20){\line(1,0){20}}
\put(25,8){\large$b$}
\put(40,20){\line(1,1){90}}
\put(40,110){\line(1,-1){90}}
\setlength\fboxsep{0pt}
\put(55,63){\colorbox{black}{\framebox(60,4){}}}

\put(130,20){\line(1,0){15}}
\put(130,110){\line(1,0){15}}

\put(145,5){\framebox(30,30){\large$\varphi_2$}}
\put(145,95){\framebox(30,30){\large$\varphi_1$}}

\put(175,20){\line(1,0){20}}
\put(175,110){\line(1,0){20}}

\put(195,5){\dashbox{4}(35, 120){\shortstack{\large D\\\large E\\\large T\\\large E\\\large C\\\large T\\\large I\\\large O\\\large N}}}
\end{picture}
\caption{Two modes $a$ and $b$ are incident on a 50:50 beam splitter.  After the beam splitter, phase shifts
$\varphi_1$ and $\varphi_2$ are imposed in the two arms.  A measurement is then made to detect
the differential phase shift $\phi_d=\varphi_1-\varphi_2$.  When the measurement is pushed beyond a second 50:50 beam splitter, i.e., when the DETECTION box includes a 50:50 beam splitter before measurement, the result is a Mach-Zehnder interferometer, which is sensitive only to $\phi_d$.}
\label{setting}
\end{center}
\end{figure}

In this brief sketch, we consider an interferometric setting, depicted in Fig.~\ref{setting}, for determining the differential phase shift imparted to fields in the interferometer's two arms.  Two modes, with annihilation operators $a$ and $b$, are incident on a 50:50 beam splitter; after the beam splitter, the two arms experience phase shifts $\varphi_1$ and $\varphi_2$.  To make the depicted setup into an interferometer, one would add a second 50:50 beam splitter, at which the modes in the two arms are recombined.  A prime reason for using such an interferometer is that it is insensitive to common-mode noise in the two arms, with each arm acting as a phase reference for the other; the interferometer is sensitive only to the differential phase shift $\phi_d=\varphi_1-\varphi_2$.  In the following, we perform a quantum Fisher analysis to determine the optimal sensitivity for estimating $\phi_d$; this analysis reports the optimal sensitivity without having to consider the second beam splitter in the interferometer.

We assume that the state input to the beam splitter is a product state~$\ket{\psi_{\rm in}}= \ket{\xi} \otimes \ket{\chi}$, where $\ket\xi$ is the state of mode~$a$ and $\ket\chi$ is the state of mode~$b$.  The action of the beam splitter is described by the unitary operator $B=e^{-i (a^{\dagger} b + b^{\dagger} a) \pi /4}$, so the state after the beam splitter is $B\ket{\psi_{\rm in}}$.  The phase shifters are described by the unitary operator  $U=e^{i(\varphi_1a^\dagger a+\varphi_2b^\dagger b)}$.  The state after the phase shifters is thus
\begin{align}
\ket{\psi} = U B\ket{\psi_{\rm in}}
\end{align}
This setup is very close to the setting we considered in~\cite{Lang2013}, the physical difference being that in~\cite{Lang2013}, the input state of mode~$a$ was required to be a coherent state.  We use the same notation here as in~\cite{Lang2013}, except that there modes $a$ and $b$ were called $a_1$ and $a_2$.

Our restriction to product states input to the initial beam splitter is natural---indeed, it is the only sensible assumption---in the case of an interferometric setup.  Product inputs do generally lead to modal entanglement, i.e., entanglement between the two arms, after the initial beam splitter.  In an interferometric setup, one is relying on the beam splitter to create modal entanglement from product inputs.  If, in contrast, one allowed arbitrary, entangled states $\ket{\Xi}$ of the two modes to be incident on the beam splitter, one could dispense with the initial beam splitter, since one could just as well input any entangled state $B\ket{\Xi}$ directly into the two arms approaching the phase shifters.

To quantify the sensitivity of a particular product input, we use quantum Fisher information.  Since we are interested in the differential phase shift, the relevant element of the Fisher matrix, called $\sF_{dd}$ in~\cite{Lang2013}, is
\begin{align}\label{eq:F1}
\begin{split}
\sF
&=\langle\psi_{\rm in}|B^\dagger N_d^2 B|\psi_{\rm in}\rangle
-\langle\psi_{\rm in}|B^\dagger N_d B|\psi_{\rm in}\rangle^2\\
&=-\ex{(a^{\dag} b - b^{\dag} a)^2}+\ex{\da{a} b-\da{b} a}^2\\
&=-\ex{\da{a}\da{a}bb}-\ex{aa\da{b}\da{b}}+\ex{\da{a}ab\da{b}}+\ex{a\da{a}\da{b}b}\\
&\qquad+\ex{\da{a}b}^2+\ex{a\da{b}}^2-2\ex{\da{a}b}\ex{a\da{b}}
\,.
\end{split}
\end{align}
Here we introduce a notation that we use throughout the following: $\langle O \rangle = \langle\psi_{\rm in}|O|\psi_{\rm in}\rangle$ denotes an expectation value with respect to the input state $\ket{\psi_{\rm in}}$.  Notice that the Fisher information is the variance of $N_d$ in the state $B|\psi_{\rm in}\rangle$ after the 50:50 beam splitter.

To find the optimal performance, we maximize $\sF$ over all product input states, subject to whatever additional constraints we impose on the input; i.e., we find the input product state that maximizes the variance of $N_d$ after the first beam splitter.  The expression~(\ref{eq:F1}) is valid for arbitrary inputs; specializing to product inputs gives
\begin{align}
\begin{split}
\sF
&=2 N_a N_b + N_a+ N_b - \ex{\da{a}\da{a}} \ex{bb}- \ex{aa} \ex{\da{b}\da{b}}\\
&\qquad-2|\!\ex{a}\!|^2\,|\!\ex{b}\!|^2
+\ex{\da{a}}^2\ex{b}^2+\ex{a}^2 \ex{\da{b}}^2\,,
\label{eq:F}
\end{split}
\end{align}
where $N_a=\ex{\da{a}a}$ and $N_b=\ex{\da{b}b}$ are the mean photon numbers in the two input modes.

Liu {\it et al.} \cite{Liu2013} have considered a setup similar to the one we consider here, a Mach-Zehnder interferometer with a product-state input. They focused on the Fisher information for an arbitrary state in mode~$a$ and a state that is a superposition of even or odd photon numbers in mode~$b$.

In the remainder of the paper, we first find the optimal input state for a fixed photon number and then find the optimal state for a constraint on mean photon number.

\section{Fixed photon number}

If we fix the total photon number $N= N_a+N_b$, all product states have the form
$|n\rangle \otimes |N-n\rangle$.  Under these circumstances, only the first three terms in Eq.~(\ref{eq:F}) contribute to the Fisher information.  Finding the maximum reduces to finding the $n$ that maximizes $2n(N-n)$; the maximum is achieved at $n=N/2$ for $N$ even and at $n=(N\pm1)/2$ for $N$ odd, the two signs corresponding to an exchange of the input modes.  The maximal Fisher information is
\begin{equation}\label{eq:FmaxN}
\sF_{\rm max}=
\begin{cases}
\displaystyle{\frac{N(N+2)}{2}}\,,&\mbox{$N$ even,}\\[10pt]
\displaystyle{\frac{N(N+2)-1}{2}}\,,&\mbox{$N$ odd.}
\end{cases}
\end{equation}
The optimal state,
\begin{equation}\label{eq:optpsiN}
|\psi_{\rm in}\rangle_N=
\begin{cases}
|N/2\rangle\otimes|N/2\rangle\,,&\mbox{$N$ even,}\\[4pt]
|(N\pm1)/2\rangle\otimes|(N\mp1)/2\rangle\,,&\mbox{$N$ odd.}
\end{cases}
\end{equation}
is the \emph{twin-Fock state\/} for $N$ even~\cite{Holland1993} and its closest equivalent for $N$ odd.  For brevity, we use ``twin-Fock state'' to refer to both the even and odd input states in the following; when we need to distinguish even and odd $N$, we refer to the former as ``identical twins'' and the latter as ``fraternal twins.''

The optimal state gives rise to a Quantum Cram{\'e}r-Rao Bound (QCRB) for the variance of the phase estimate given by
\begin{align}
(\Delta\phi_d^{\rm est})^2\ge\frac{1}{\sF_{\rm max}}=
\begin{cases}
\displaystyle{\frac{2}{N(N+2)}}\,,&\mbox{$N$ even,}\\[10pt]
\displaystyle{\frac{2}{N(N+2)-1}}\,,&\mbox{$N$ odd,}
\end{cases}\label{eq:cases}
\end{align}
which shows an asymptotic Heisenberg scaling.  An input twin-Fock state leads to modal entanglement between the two arms after the beam splitter~\cite{Zhang2013}.

Holland and Burnett~\cite{Holland1993} introduced the twin-Fock state (for $N$ even) and considered the Heisenberg scaling of its phase sensitivity.  Measurements that achieved the Heisenberg scaling were demonstrated in~\cite{Kim98,Campos03,Gerry10}.  Robustness of the identical-twin-Fock state against various errors was investigated in~\cite{Kim98} and~\cite{Meiser09}, and sub-shot-noise precision for interferometry with identical-twin-Fock states was demonstrated experimentally in~\cite{Sun08}.

Benatti {\it et al.}~\cite{Benatti2010} considered the Fisher information for detecting a differential phase shift in an interferometric setting, with the constraint that there be no entanglement between the two inputs to the interferometer.  They showed that the identical-twin-Fock state has the Fisher information expressed by the $N$-even case of Eq.~(\ref{eq:FmaxN}).

If we were to remove the restriction of having a product input to the interferometer, the optimal input would be the state that maximizes the variance of $N_d$ after the first beam splitter, i.e. the modally entangled input state $|\Xi\rangle$ that becomes a \emph{N00N state}, $B|\Xi\rangle=(|N,0\rangle+|0,N\rangle)/\sqrt2$, after the beam splitter~\cite{Gerry00,Boto00,Dowling08}.  The N00N state has a Fisher information $\sF=N^2$; this is generally larger than the Fisher information~(\ref{eq:FmaxN}) of the twin-Fock input, because one is optimizing over a larger set of input states to the initial beam splitter.  For $N=1$ and $N=2$, however, the twin-Fock product input does produce a N00N-like state on the other side of the beam splitter: the $N=1$ fraternal-twin-Fock input, $|1\rangle\otimes|0\rangle$, leads to the N00N-like state $B|1,0\rangle=(|1,0\rangle-i|0,1\rangle)/\sqrt2$ after the beam splitter, and the $N=2$ identical-twin-Fock input, $|1\rangle\otimes|1\rangle$, leads to the N00N state $B|1,1\rangle=-i(|2,0\rangle+|0,2\rangle)/\sqrt2$.  Thus the twin-Fock inputs for $N=1$ and $N=2$ have the same Fisher information as the N00N state; for $N>2$, however, the N00N-state Fisher exceeds that of the twin-Fock input and is a factor of 2 larger asymptotically for large~$N$.

\section{Fixed mean photon number}

We now move on to a constraint on the mean photon number $N_a+N_b$; in this section $N$ denotes this total mean photon number.  The proof for the optimal input state consists of two steps.  The first step finds the optimal states $\ket{\xi}$ and $\ket{\chi}$ under the assumption that both $N_a$ and $N_b$ have fixed values.  It turns out that the form of the optimal states is independent of how the total mean photon number $N$ is divided up between $N_a$ and $N_b$.  In the second step, we show that the optimal split of resources is an equal division, $N_a=N_b=N/2$.

We begin the first step  by noticing that the quantity on the second line of Eq.~(\ref{eq:F}) is either negative or zero.  We ignore this quantity for the moment.  As we see shortly, the product input state that maximizes the top line has $\ex{a}=\ex{b}=0$ and, therefore, also maximizes the quantum Fisher information~$\sF$.  Furthermore, in this first step, $N_a$ and $N_b$ are assumed to be fixed, so maximizing the top line reduces to maximizing
\begin{align}
 - \ex{\da{a}\da{a}}\ex{bb}-\ex{aa}\ex{\da{b}\da{b}}.\label{eq:part}
\end{align}
We can always choose the phase of mode~$a$, i.e., multiply~$a$ by a phase factor, to make $\ex{aa}$ real and positive, i.e., $\ex{aa}=\ex{\da{a}\da{a}}\ge0$.  With this choice we need to maximize
\begin{align}
-\ex{aa}(\ex{bb}+\ex{\da{b}\da{b}})=\ex{aa}(\langle p^2\rangle-\langle x^2\rangle)\,,
\label{eq:bmax}
\end{align}
where in the second form we introduce the quadrature components $x$ and $p$ for mode~$b$, i.e., $b=(x+ip)/\sqrt{2}$.

The proof continues along the lines of~\cite{Lang2013}:
\begin{align}
\begin{split}
\left( \langle p^2\rangle - \langle x^2\rangle \right)^2
&=\left( \langle p^2\rangle + \langle x^2\rangle \right)^2 - 4\langle x^2\rangle \langle p^2\rangle \\
&= \left(2 N_b +1 \right)^2 - 4\langle x^2\rangle \langle p^2\rangle \\
&\le \left(2 N_b +1 \right)^2 - 4\langle (\Delta x)^2\rangle \langle (\Delta p)^2\rangle \\
&\le \left(2 N_b +1 \right)^2 - 1 \\
&= 4 N_b(N_b+1)\,.
\end{split}
\end{align}
The first inequality is saturated if and only if $\ex{x}=\ex{p}=0$; equality is achieved in the second inequality if and only if the input state $\ket{\chi}$ of mode~$b$ is a minimum-uncertainty state.  This situation is identical to that in~\cite{Lang2013}: the optimal choice for $\ket{\chi}$ is squeezed vacuum with $x$ the squeezed quadrature and $p$ the anti-squeezed quadrature.

What is left now is to maximize
\begin{align}
2\sqrt{N_b(N_b+1)}\ex{aa}=\sqrt{N_b(N_b+1)}(\ex{aa}+\ex{\da{a}\da{a}})
\end{align}
over the input states $\ket\xi$ of mode~$a$ for which $\ex{aa}$ is real and positive.  This is the same maximization we just performed for mode~$b$, except for a sign change, whose effect is to exchange the squeezed and anti-squeezed quadratures.  The optimal state $\ket\xi$ is squeezed vacuum with $p$ as the squeezed quadrature and $x$ as the anti-squeezed quadrature.

Summarizing, the optimal input state is $S_a(-r)|0\rangle\otimes S_b(r')|0\rangle$, where $r$ and $r'$ are real and positive and $S_c(\gamma)=\exp[\frac12(\gamma c^2 -\gamma^* c^{\dagger 2})]$ is the squeeze operator for a field mode $c$.  The values of the squeeze parameters are determined by $N_a=\sinh^2 r$ and $N_b=\sinh^2 r'$.  Notice that, as promised, the optimal state has $\ex{a}=\ex{b}=0$ and thus maximizes the Fisher information~(\ref{eq:F}); the maximum value is
\begin{align}
\sF=2N_a N_b+N_a+N_b+2\sqrt{N_a(N_a+1)N_b(N_b+1)}\,.
\label{eq:sepnumbers}
\end{align}

The second step of the proof is now trivial.  For a constraint on the total mean photon number $N=N_a+N_b$, it is straightforward to see that Eq.~(\ref{eq:sepnumbers}) is maximized by splitting the photons equally between the two modes, i.e., $N_a=N_b=N/2$.  The resulting optimal input state has $r=r'$,
\begin{align}
|\psi_{\rm in}\rangle_{\rm opt}
=S_a(-r)|0\rangle\otimes S_b(r)|0\rangle\,,
\label{eq:optstate}
\end{align}
and the maximal Fisher information and corresponding QCRB are
\begin{align}
\sF= N(N+2)\,,\quad
(\Delta\phi_d^{\rm est})^2\ge\frac{1}{\sF_{\rm max}}=\frac{1}{N(N+2)}\,.
\label{eq:FmaxmeanN}
\end{align}
This again exhibits Heisenberg scaling and, without the factor of 2 that appears in the fixed-photon-number result~(\ref{eq:FmaxN}), achieves the $1/N^2$ Heisenberg limit.

A question that naturally arises is that of the optimal measurement.  Here again we can refer to~\cite{Lang2013}, which showed in the Supplemental Material, building on work of Pezze and Smerzi~\cite{Pezze08}, that the classical Fisher information of photon counting after a second 50:50 beam splitter is the same as the quantum Fisher information, provided the coefficients of the expansion of the input state in the number basis are real.  This requirement is met by the optimal state~(\ref{eq:optstate}).

Unlike the situation where there is a strong mean field, however, the interferometer with dual squeezed-vacuum inputs runs on modulated noise, so the mean of the differenced photocount after a second 50:50 beam splitter gives no information about the phase.  One strategy for extracting the phase information is to look directly at the fluctuations by squaring the differenced photocount and thus effectively measuring $N_d^2$~\cite{Kim98,Tonekaboni14}; one can show~\cite{footnote1} that the sensitivity at the optimal operating point achieves the QCRB~(\ref{eq:FmaxmeanN}).

Notice now that since $BaB^\dagger=(a+ib)/\sqrt2$ and $BbB^\dagger=(b+ia)/\sqrt2$, we have $B(a^2-b^2)B^\dagger=a^2-b^2$.  Thus the beam splitter leaves unchanged the product of squeeze operators in the optimal input state~(\ref{eq:optstate}),
\begin{equation}\label{eq:Bsqueeze}
BS_a(-r)S_b(r)B^\dagger=S_a(-r)S_b(r)\,,
\end{equation}
and this in turn means that the optimal input state is an eigenstate of the beam splitter,
\begin{align}
B|\psi_{\rm in}\rangle_{\rm opt}=
BS_a(-r)S_b(r)|0,0\rangle=|\psi_{\rm in}\rangle_{\rm opt}\,.
\label{eq:Bpsi}
\end{align}
Thus the state after the 50:50 beam splitter is the same product of squeezed vacua as before the beam splitter~\cite{twomode}.  The Heisenberg limit is thus achieved without any entanglement between the arms of the interferometer.  In fact, Jiang, Lang, and Caves~\cite{Zhang2013} showed that the state $|\psi_{\rm in}\rangle_{\rm opt}$ is the only nonclassical product state, i.e., not a coherent state, that produces no modal entanglement after a beam splitter.  These results indicate that, as in~\cite{Sahota14}, modal entanglement is not a crucial resource for quantum-enhanced interferometry.

Caves pointed out that using squeezed states in an interferometer allows one to achieve sensitivities below the shot-noise limit~\cite{Caves81}; this original scheme, often simply dubbed ``squeezed-state interferometry,'' involves injecting squeezed vacuum into the secondary input port of an interferometer.  That squeezing the light into the primary input port, in addition to inputting squeezed light into the secondary port, is advantageous was first shown by Bondurant and Shapiro~\cite{Bondurant84} and further investigated by Kim and Sanders~\cite{Kim96}.   All these papers, however, included a mean field in at least one of the input modes.  Paris argued~\cite{Paris99} that if one considers arbitrary squeezed-coherent states as interferometer inputs, putting all the available power into the squeezing, instead of into a mean field, yields better fringe visibility.  Under a Gaussian constraint, Refs.~\cite{Pinel2012} and~\cite{Demkowicz2014} showed that a state that maximizes the Fisher information for a detecting a differential phase shift after a beam splitter is dual squeezed vacua; relative to these last results, our contribution in this paper is to remove the assumption of Gaussianity, replacing it with a restriction to product inputs.

A problem with using Fisher information to find optimal states under a mean-number constraint is that one can come up with states that seemingly violate the Heisenberg limit.  This was noted for single-mode schemes by Shapiro~\cite{Shapiro89} and later by Rivas~\cite{Rivas2012}. For the former case, Braunstein and co-workers showed that under a precise asymptotic analysis, no violation of the Heisenberg limit occurs~\cite{Braunstein1992b,Braunstein1992c,Lane1993a}.  For the latter case, it was shown that the Fisher information does not provide a tight bound, which makes a Fisher analysis uninformative\cite{Tsang2012,Giovannetti12}.  If we were to allow arbitrary (entangled) states $|\Xi\rangle$ as inputs in our scheme, we would run into the same problem~\cite{footnote2}.  Requiring product inputs removes this pathology of the Fisher information, therefore providing additional motivation for our product constraint.

\acknowledgments
The authors thank S.~Szigeti and S.~Haine for useful discussions.  This work was supported in part by NSF
Grant Nos.~PHY-1212445 and PHY-1314763.


\begin{thebibliography}{99}
\bibitem{footnote1}
The squared error in estimating $\phi _d$ from $\langle N_d^2\rangle$ is given by
$(\Delta\phi^{\rm est}_d)^2=\mathop{\protect\rm Var}(N_d^2)/|\partial\langle N_d^2\rangle/\partial_{\phi_d}|
=1/N(N+2) + (\cos^2\phi_d/4\sin^2\phi_d)[7+2/N+N/(N+2)]$, from which we can see that at the optimal operating point, $\phi _d\simeq\pi/2$, this procedure achieves the QCRB.
\bibitem{twomode}
Since $B(a^2+b^2)B^\dagger=2iab$, we have $BS_a(r)S_b(r)B^\dagger=e^{ir(ab+a^\dagger b^\dagger)}$.  This means that dual, equally squeezed vacua into the two input ports, $S_a(r)|0\rangle\otimes Sb(r)|0\rangle$, become a two-mode squeezed state $e^{ir(ab+a^\dagger b^\dagger)}|0\rangle\otimes|0\rangle$ after the beam splitter.  This well-known situation is not, however, the one that gives maximal Fisher information.
\bibitem{footnote2}
Since the Fisher information is the variance of $N_d$ after the beam splitter, it is given by $\mathop {\protect \rm  Var}(N_d)={\protect\rm  Var}(N_a)+{\protect \rm  Var}(N_b) - 2 \mathop {\protect\rm  Covar}(N_a,N_b)$.  If mode~$b$ is in vacuum, we have ${\protect\rm  Var}(N_d)={\protect\rm  Var}(N_a)$, and this is the quantity that was made arbitrarily large under a mean-number constraint in~\cite{Rivas2012}.
\end{thebibliography}
\end{document}